\documentclass[reprint,amsmath,amssymb,aps,superscriptaddress,floatfix]{revtex4-1}

\usepackage{graphicx}
\usepackage{dcolumn}
\usepackage{bm}
\usepackage{hyperref}
\usepackage{siunitx}
\usepackage{color}
\usepackage{xspace}

\DeclareSIUnit\kT{$k_B T$}
\DeclareSIUnit\dyne{dyne}
\DeclareSIUnit\molar{M}
\newcommand{\bnabla}{\mbox{\boldmath$\nabla$}}
\newcommand{\bdiv}{\bnabla\cdot}
\newcommand{\bcurl}{\bnabla\times}
\newcommand{\dd}{\mathrm{d}}
\newcommand{\ee}{\mathrm{e}}

\newcommand{\ve}[1]{\bm{\mathbf{#1}}}
\newcommand{\veh}[1]{\bm{\mathbf{\hat{#1}}}}

\newcommand{\kT}{{k_B T}}

\newcommand{\nonum}{\nonumber\\}

\newcommand{\rmv}{\textrm{v}}
\newcommand{\rmt}{\textrm{t}}
\newcommand{\itfd}{\textit{fd}}
\newcommand{\rmMK}{\textrm{M13}}
\newcommand{\fd}{\textit{fd}\xspace}
\newcommand{\MK}{M13\xspace}

\begin{document}

\title{Chiral twist drives raft formation and organization in membranes composed of rod-like particles}

\date{\today}

\author{Louis Kang}
\email{lkang@mail.med.upenn.edu}
\affiliation{Department of Physics \& Astronomy, University of Pennsylvania, 209 South 33rd Street, Philadelphia, Pennsylvania 19104, USA}

\author{T. C. Lubensky}
\affiliation{Department of Physics \& Astronomy, University of Pennsylvania, 209 South 33rd Street, Philadelphia, Pennsylvania 19104, USA}

\begin{abstract}
Lipid rafts are hypothesized to facilitate protein interaction, tension regulation, and trafficking in biological membranes, but the mechanisms responsible for their formation and maintenance are not clear. Insights into many other condensed matter phenomena have come from colloidal systems, whose micron-scale particles mimic basic properties of atoms and molecules but permit dynamic visualization with single-particle resolution. Recently, experiments showed that bidisperse mixtures of filamentous viruses can self-assemble into colloidal monolayers with thermodynamically stable rafts exhibiting chiral structure and repulsive interactions. We quantitatively explain these observations by modeling the membrane particles as chiral liquid crystals. Chiral twist promotes the formation of finite-sized rafts and mediates a repulsion that distributes them evenly throughout the membrane. Although this system is composed of filamentous viruses whose aggregation is entropically driven by dextran depletants instead of phospholipids and cholesterol with prominent electrostatic interactions, colloidal and biological membranes share many of the same physical symmetries. Chiral twist can contribute to the behavior of both systems and may account for certain stereospecific effects observed in molecular membranes.
\end{abstract}

\maketitle

\section{Introduction}

Filamentous viruses have proven to be a fruitful colloidal system~\cite{Dogic:2000tp,Purdy:2003bj,Dogic:2004id,Lettinga:2005cw,Dogic:2006bq,Tombolato:2006hy,Barry:2009uv,Barry:2009vo,Pelcovits:2009tz,Barry:2010cz,Kaplan:2010tk,Yang:2012ds,Gibaud:2012cf,Tu:2013by,Tu:2013bg,Kaplan:2014cz,Zakhary:2014de,Sharma:2014cl,Kang:2016is}. They serve as monodisperse, rigid, and chiral rods that are approximately one micron in length and interact effectively through hard-core repulsion~\cite{Barry:2009uv,Purdy:2003bj}. When suspended in an aqueous solution at increasing concentrations, they transition from a disordered isotropic phase to a cholesteric (chiral nematic) phase characterized by alignment along a director field that twists with a preferred handedness and wavelength~\cite{Dogic:2000tp,Tombolato:2006hy}. The addition of a non-adsorbing polymer such as dextran induces lateral virus-virus attraction via the depletion interaction~\cite{Asakura:1954jy,Asakura:1958kc,Barry:2010cz,Yang:2012ds}. The viruses self-assemble into monolayers that exhibit fluid-like dynamics internally~\cite{Barry:2010cz} and sediment to the bottom of glass containers, which are coated with a polyacrylamide brush to suppress depletion-induced virus-wall attractions~\cite{Lau:2009gw}. The rich physics and phenomenology of membranes formed from single virus species have been thoroughly studied~\cite{Barry:2009vo,Pelcovits:2009tz,Barry:2010cz,Kaplan:2010tk,Yang:2012ds,Gibaud:2012cf,Tu:2013by,Tu:2013bg,Kaplan:2014cz,Zakhary:2014de,Kang:2016is}. However, two-species membranes demonstrate a novel set of behaviors which are not adequately understood~\cite{Sharma:2014cl}. We will review these behaviors now before describing a theory that can explain them.

\begin{figure*}
	\includegraphics[width=\linewidth]{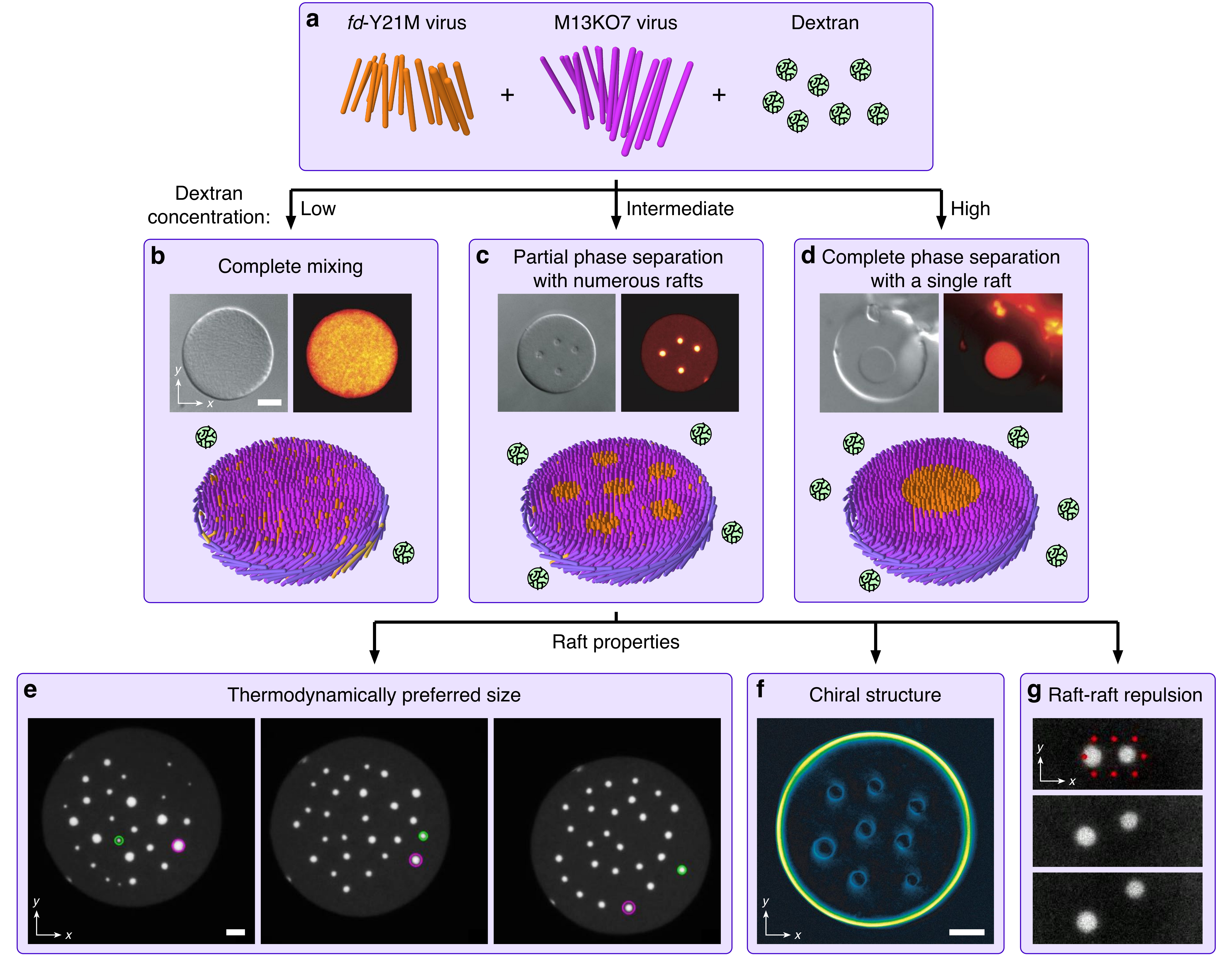}
	\caption{\label{fig:overview}Overview of two-species colloidal membrane experiments. \textbf{a}, Virus particles and dextran molecules act as rod-shaped colloids and spherical depletants, respectively. \fd viruses are shorter and prefer right-handed twist. \MK viruses are longer and prefer left-handed twist. \textbf{b}--\textbf{d}, Differential interference contrast image (top left), fluorescence image with \fd labeled (top right), and schematic (bottom) of colloidal membranes. \textbf{b}, At a low dextran concentration of $\SI{41000}{\per\cubic\um}$, the two virus species completely mix. \textbf{c}, At an intermediate dextran concentration of $\SI{46000}{\per\cubic\um}$, several smaller rafts of \fd virus form in a partially phase-separated background. \textbf{d}, At a high dextran concentration of $\SI{62000}{\per\cubic\um}$, the two virus species completely phase separate. \textbf{e}, Rafts exchange rods with the background membrane to attain a thermodynamically preferred size. Fluorescence images with \fd labeled taken \SI{6.7}{\hour} apart. Green and purple circles track two rafts that start, respectively, smaller and larger than the preferred raft size. \textbf{f}, Viruses adopt a twisted chiral structure. LC-PolScope birefringence map with pixel brightness representing retardance, which indicates virus tilt toward the membrane plane. \textbf{g}, Rafts repel one another. Fluorescence images with \fd labeled taken \SI{5}{\s} apart. Two optical plows consisting of multiple light beams (red dots) bring two rafts together and are then switched off. All scale bars, \SI{5}{\um}. Experimental data and methods are reported in Ref.~\cite{Sharma:2014cl}. Schematics not drawn to scale. Microscopy images reproduced with permission from Nature Publishing Group.}
\end{figure*}

\begin{table*}
	\caption{\label{tab:raft}Membrane parameters and their values.}
	\begin{ruledtabular}\begin{tabular}{ccccc}
		Parameter & Variable & Experimental estimate & Reference(s) & Model value \\
		\hline
		\textit{fd}-Y21M half-length & $l_\itfd$ & $\SI{430}{\nm}$ & \cite{Sharma:2014cl}\footnotemark[1] & same \\
		M13KO7 half-length & $l_\rmMK$ & $\SI{560}{\nm}$ & \cite{Sharma:2014cl}\footnotemark[1] & same \\
		Virus half-length difference & $d$ & $\SI{130}{\nm}$ & $l_\rmMK - l_\itfd$ & same \\
		Virus diameter & & $\SI{7}{\nm}$ & \cite{Sharma:2014cl} & \\
		Virus nearest-neighbor distance & $\xi$ & $\SI{12}{\nm}$ & \cite{Kang:2016is} & same \\
		Virus 2D concentration & $c_\textrm{v}$ & ${\sim}\SI{9000}{\per\square\um}$ & $1/\pi(\xi/2)^2$ & $\SI{8500}{\per\square\um}$ \\
		\textit{fd}-Y21M Frank constant & $K_\itfd$ & ${\sim}\SI{2}{\pico\N}$ & \cite{Dogic:2000tp}\footnotemark[2]\footnotemark[3] & \SI{4}{\pico\N} \\
		M13KO7 Frank constant & $K_\rmMK$ & ${\sim}\SI{4}{\pico\N}$ & \cite{Tombolato:2006hy}\footnotemark[3] & \SI{10}{\pico\N} \\
		\textit{fd}-Y21M twist wavenumber & $q_\itfd$ & ${\sim}\SI{0.1}{\per\um}$ & \cite{Barry:2009uv}\footnotemark[3] & \SI{0.11}{\per\um} \\
		M13KO7 twist wavenumber & $q_\rmMK$ & ${\sim}\SI{-0.5}{\per\um}$ & \cite{Tombolato:2006hy}\footnotemark[3] & \SI{-0.55}{\per\um} \\
		\textit{fd}-Y21M birefringence & $\Delta n_\itfd$ & ${\sim}0.008$ & \cite{Barry:2009vo}\footnotemark[2]\footnotemark[4] & $0.011$ \\
		M13KO7 birefringence & $\Delta n_\rmMK$ & ${\sim}0.008$ & \cite{Barry:2009vo}\footnotemark[2]\footnotemark[4] & $0.011$ \\
		Dextran concentration & $c$ & \SI{48000}{\per\cubic\um} & \cite{Sharma:2014cl} & same \\
		Dextran radius & $a$ & ${\sim}\SI{25}{\nm}$ & \cite{Ioan:2000kl,Armstrong:2004kh,Banks:2005cc}\footnotemark[5] & same \\
		Temperature & $T$ & \SI{22}{\celsius} & \cite{Sharma:2014cl} & same \\
	\end{tabular}\end{ruledtabular}
	\footnotetext[1]{Half the end-to-end length estimated from contour lengths and persistence lengths.}
	\footnotetext[2]{Measured for \itfd-wt virus.}
	\footnotetext[3]{Imprecise estimates extrapolated to membrane virus concentration ${\sim}$\SI{200}{\mg\per\mL} (corresponding to $c_\rmv \sim \SI{9000}{\per\square\um}$) based on concentration-dependent behavior of \itfd-wt suspensions~\cite{Dogic:2000tp}.}
	\footnotetext[4]{Assuming membrane nematic order parameter of 1 and virus concentration ${\sim}$\SI{200}{\mg\per\mL} (corresponding to $c_\rmv \sim \SI{9000}{\per\square\um}$).}
	\footnotetext[5]{Hydrodynamic radii for dilute solutions of \SI{500}{\kilo\dalton} dextran, whereas our experiments are in the semidilute regime.}
\end{table*}

\textit{fd}-Y21M and M13KO7, which we will shorten to \fd and \MK for convenience, are two species of filamentous virus that have slightly different lengths and form cholesteric phases of opposite handednesses (Table~\ref{tab:raft} and Fig.~\ref{fig:overview}a). Membranes composed of both \fd and \MK viruses are circular with interior particles aligned largely perpendicularly to the membrane plane and edge particles tilted azimuthally, as in single-species membranes~\cite{Gibaud:2012cf}. At low dextran concentrations, the two species are fully mixed, and at high dextran concentrations, the two are fully phase-separated with \MK viruses surrounding a single \fd domain (Fig.~\ref{fig:overview}b,d). At intermediate concentrations, membranes exhibit partial phase separation with several smaller circular rafts of \fd viruses distributed within a mixed background of both species (Fig.~\ref{fig:overview}c).

Particle tracking experiments show that \fd viruses diffuse in and out of these rafts~\cite{Sharma:2014cl}, allowing for equilibration to a thermodynamically preferred raft size over ${\sim}\SI{24}{\hour}$ (Fig.~\ref{fig:overview}e). Polarized light microscopy suggests that the raft system has a chiral structure, with particles tilting around the interfaces between rafts and background membrane and around the membrane edge (Fig.~\ref{fig:overview}f). Finally, the rafts are distributed homogeneously throughout the membrane and never coalesce, indicating a long-ranged repulsion between rafts (Fig.~\ref{fig:overview}g). This interaction can be measured quantitatively by bringing two rafts close together with optical traps and tracking their trajectories upon release of the traps~\cite{Sharma:2014cl}.

The simplicity of this colloidal membrane system allows us to study it theoretically with a model built from established physical principles and experimentally meaningful parameters. Its components have well-characterized interactions: dextran molecules act as depletants that interact with viruses through hard-body interactions~\cite{Asakura:1954jy,Asakura:1958kc,Barry:2010cz,Yang:2012ds,harnaudietrich}, and the hard-body interactions between viruses can be coarse-grained as the Frank free energy for chiral liquid crystals~\cite{Dogic:2000tp,Tombolato:2006hy}. We previously used such a model to investigate single-species membranes and succeeded in reproducing a variety of structural, dynamical, and phase phenomena with a single set of realistic parameter values~\cite{Kang:2016is}. Extending the model to the two-species system will demonstrate how the intruiging behaviors depicted in Fig.~\ref{fig:overview} emerge from Frank free energy, depletant entropy, and mixing entropy.

The fundamental principles we encounter on the colloidal scale may apply to similar but less tractable molecular systems whose particles and interactions share the same physical symmetries. Colloidal systems have permitted the investigation of many quintessential condensed matter phenomena with single-particle resolution and exquisite control. For example, spherical colloids exhibit crystal nucleation~\cite{Gast:1983if,Gasser:2001fo} and glassy dynamics~\cite{Pusey:1986dv,Weeks:2000dw}; the addition of an isotropic attraction with depletants allows them to demonstrate liquid-gas phase separation~\cite{Lekkerkerker:1992ij}, thermal capillary waves~\cite{Aarts:2004hi}, and wetting~\cite{Aarts:2004go}. And in addition to the aforementioned work in which filamentous viruses form nematic and cholesteric liquid crystal phases, plate-like and rod-like colloids have shed insight on columnar and smectic liquid crystal phases, respectively~\cite{Lekkerkerker:2000ct,Lettinga:2007fe}. Phospholipid fluid membranes are another important soft-matter system; yet, due to our inability to directly visualize real-time dynamics of lipid bilayers at the nanometer scale, many processes remain poorly understood. Following the analogy between colloids and molecular substances, our theoretical investigation of two-component colloidal membranes may provide new, universal understanding about membrane rafts, which have been observed in experimental phospholipid membranes~\cite{Dietrich:2001jm,Veatch:2003hn} but remain controversial in the case of biological membranes~\cite{Lingwood:2009kf}.

As shown in previous work~\cite{Xie:2016gv,sakhardande} based on phenomenological models, the difference in chirality between two coexisting phases, which favors different twist rates of viruses relative to membrane normals, is the primary driver of raft formation in viral membranes. When two achiral phases coexist, the interface separating them has a positive line tension (or surface tension in three dimensions) that favors the smallest possible interfacial length (or area). Chirality difference introduces an effective negative contribution to the line tension, which for large enough difference becomes negative and favors as much interfacial length as possible. Finite-size rafts are a result of the competition between negative line tension and either repulsive interaction between segments of interface or interfacial curvature energy. The repulsive energy between rafts as they approach each other arises from compression of the twist in the membranes' background phase. The formation of rafts and their mutual interaction follows this fundamental physics in our calculations that are based on the particular depletion physics of viral rafts.

The next few sections describe, respectively, the process of phase separation that generates raft and background phases, the organization of the raft phase into domains with a preferred size and chiral structure, and the repulsion between rafts mediated by the chiral structure of the background phase. Each section includes theoretical development, results, and comparison to experimental data. In the last section, we discuss the assumptions made by our theory, its contribution to the literature on heterogeneous membranes, and implications for phospholipid membranes.

\section{Phase separation between virus species}

\begin{figure}
	\includegraphics[width=\columnwidth]{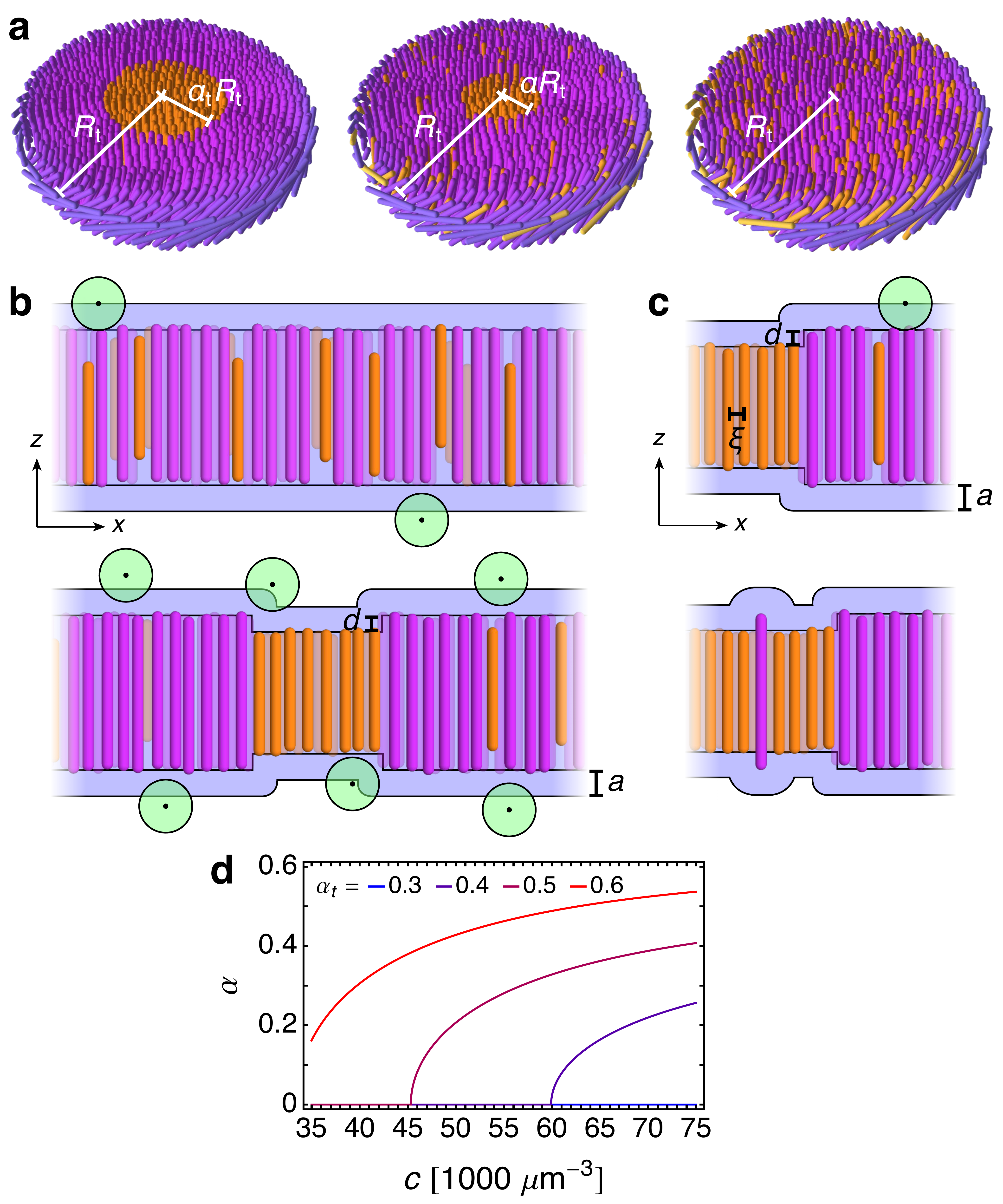}
	\caption{\label{fig:mixing}Phase separation into a raft phase containing only \fd virus (orange) and a background phase containing both \fd and \MK (purple) viruses. \textbf{a}, For a completely phase-separated membrane (left), the area fraction of the raft phase is $\alpha^2 = \alpha_\rmt^2$; equivalently, if the raft phase formed a single circular domain as depicted, it would have radius $\alpha_\rmt R_t$. As \fd viruses enter into the \MK-rich phase (middle), the area fraction of the raft phase decreases to $\alpha^2 < \alpha_\rmt^2$. For a completely mixed membrane (right), $\alpha^2 = 0$. \textbf{b}, Competition between the entropy of mixing and depletant entropy determines $\alpha$. At low depletant concentration (top), the mixed state is entropically preferred. Phase separation reduces the excluded volume and is preferred at high depletant concentration (bottom). Green circles represent depletants and blue regions represent the excluded volume. \textbf{c}, Introducing a shorter virus into a sea of longer ones (top) increases the excluded volume less than introducing a longer virus into a sea of shorter ones (bottom). \textbf{d}, $\alpha$ for various $\alpha_t$ and depletant concentrations $c$ (Eq.~\ref{eqn:alpha}). Values for other parameters are provided in Table~\ref{tab:raft}. Schematics not drawn to scale.}
\end{figure}

We start by investigating the separation of membrane particles into two phases, one which we call the ``background'' phase containing mostly \MK viruses completely surrounding the other which we call the ``raft'' phase containing \fd viruses, in accordance with experiment (Fig.~\ref{fig:overview}). The structure of the phases, including the number and size of rafts present, does not yet concern us. We assume a large circular membrane of radius $R_\rmt \rightarrow \infty$ and henceforth ignore effects of the outer boundary. The degree of phase separation is parametrized by $\alpha^2$, the area fraction of the raft phase (Fig.~\ref{fig:mixing}a). It ranges between $\alpha^2 = 0$, which corresponds to complete mixing, and $\alpha^2 = \alpha_\rmt^2$, which corresponds to complete phase separation. $\alpha_\rmt^2$ is determined experimentally by the fraction of \fd virus provided in the initial suspension. For intermediate values of $\alpha^2$, some \fd particles leave the raft and enter the background, producing a partially mixed background phase containing both viruses.

Competition between two factors determines the degree of phase separation. Thermal forces encourage the depletants to explore as much physical space as possible. To do so, they must minimize the volume excluded to their centers of mass by the membrane, which can be accomplished by separating viruses of different lengths into different phases. A shorter \fd particle produces more excluded volume when surrounded by longer \MK particles (Fig.~\ref{fig:mixing}b). For depletant particles small compared to the dimensions of the membrane, the excluded volume is approximately $V + a A$, where $V$ is the volume of the membrane, $A$ is the surface area of the membrane, and $a$ is the depletant radius~\cite{Allendoerfer:1948da}. Their free energy is calculated via the ideal gas partition function $V_\textrm{a}^N/N!\Lambda^{3N}$ applied to $N$ depletant molecules, where $\Lambda$ is their thermal de~Broglie wavelength. The volume available to the depletants can be written as $V_\textrm{a} = V_\textrm{t} - V - a A$, where $V_\textrm{t} \gg V$ is the total volume of the virus-and-depletant suspension~\cite{HansenGoos:2007fs}. Ignoring constant terms, the depletant free energy is generically
\begin{equation}
	F_\textrm{dep} = -NT \log \frac{V_\textrm{t} - V - a A}{V_\textrm{t}} \approx c T (V + a A)
	\label{eqn:Fdepgen}
\end{equation}
where $c$ is the depletant concentration and $T$ is the temperature. We use units in which the Boltzmann constant is unity.

However, thermal forces also encourage binary fluids to adopt disordered phases in which the two species are mixed. This tendency is described quantitatively by the entropy of mixing~\cite{chaikinlubensky}. As depicted in Fig.~\ref{fig:mixing}a, the mixed background phase of total area $(1-\alpha^2) \pi R_\rmt^2$ is formed from an area $(\alpha_\rmt^2-\alpha^2) \pi R_\rmt^2$ of \fd viruses and an area of $(1-\alpha_\rmt^2) \pi R_\rmt^2$ of \MK viruses, yielding respective area fractions
\begin{equation}
	\phi_\itfd = \frac{\alpha_\rmt^2-\alpha^2}{1-\alpha^2} \quad\textrm{and}\quad \phi_\rmMK = \frac{1-\alpha_\rmt^2}{1-\alpha^2} = 1-\phi_\itfd.
\end{equation}
The entropy of mixing per particle of the background phase is
\begin{equation}
	s_\textrm{mix} = \phi_\itfd \log \phi_\itfd + \phi_\rmMK \log \phi_\rmMK.
	\label{eqn:smix}
\end{equation}
We only consider mixing in the background phase because introducing the longer \MK viruses into the raft phase is disfavored by the depletants. Their surface protrusions would be surrounded by extra excluded volume of order $da^2$ per \MK particle, unlike the smaller amount of excluded volume of order $d(\xi/2)^2$ per \fd particle required to introduce the shorter \fd viruses into the background phase (Fig.~\ref{fig:mixing}c). $d \equiv l_\rmMK - l_\itfd$ is the virus half-length difference, $a$ is the depletant radius, and $\xi$ is the nearest-neighbor virus separation (Table~\ref{tab:raft}). We thus ignore mixing in the raft phase due to these asymmetric effects of surface convexity and concavity on the depletion free energy.

Combining the mixing entropy Eq.~\ref{eqn:smix} and the depletion free energy Eq.~\ref{eqn:Fdepgen}, which respectively disfavor and favor phase separation, gives the free energy
\begin{align}
	\frac{F_\textrm{sep}}{\pi R_\rmt^2 T} &= c_\rmv\left[(1-\alpha_\rmt^2) \log\frac{1-\alpha_\rmt^2}{1-\alpha^2} + (\alpha_\rmt^2-\alpha^2) \log\frac{\alpha_\rmt^2-\alpha^2}{1-\alpha^2}\right] \nonum
	& \qquad {}+ 2 c d (\alpha_\rmt^2 - \alpha^2),
	\label{eqn:Fsep}
\end{align}
where $c$ is the 3D depletant concentration and $c_\rmv$ is the 2D virus concentration in the membrane. Minimizing $F_\textrm{sep}$ with respect to $\alpha$ produces the result
\begin{equation}
	\alpha = \begin{cases} \displaystyle \sqrt\frac{\alpha_\rmt^2-\ee^{-2c d/c_\rmv}}{1-\ee^{-2c d/c_\rmv}} & c d/c_\rmv \geq \log 1/\alpha_\rmt \\ 0 & c d/c_\rmv \leq \log 1/\alpha_\rmt. \end{cases}
	\label{eqn:alpha}
\end{equation}
where $c$ is the 3D depletant concentration, $c_\rmv$ is the 2D virus concentration in the membrane, and $d$ is the half-length difference between the two species. In Fig.~\ref{fig:mixing}d, $\alpha(c)$ is plotted for various $\alpha_\rmt$'s using values in Table~\ref{tab:raft}. For each $\alpha_\rmt$, there is complete mixing ($\alpha = 0$) below a critical depletant concentration $(c_\rmv/d) \log 1/\alpha_\rmt$. Above this critical $c$, the system partially phase-separates and approaches complete phase separation for $c \rightarrow \infty$. This behavior qualitatively agrees with experimental results in Fig.~\ref{fig:overview}b--d over the experimental range of depletant concentrations $c$.

\section{Raft organization and structure}

\begin{figure*}
	\includegraphics[width=\textwidth]{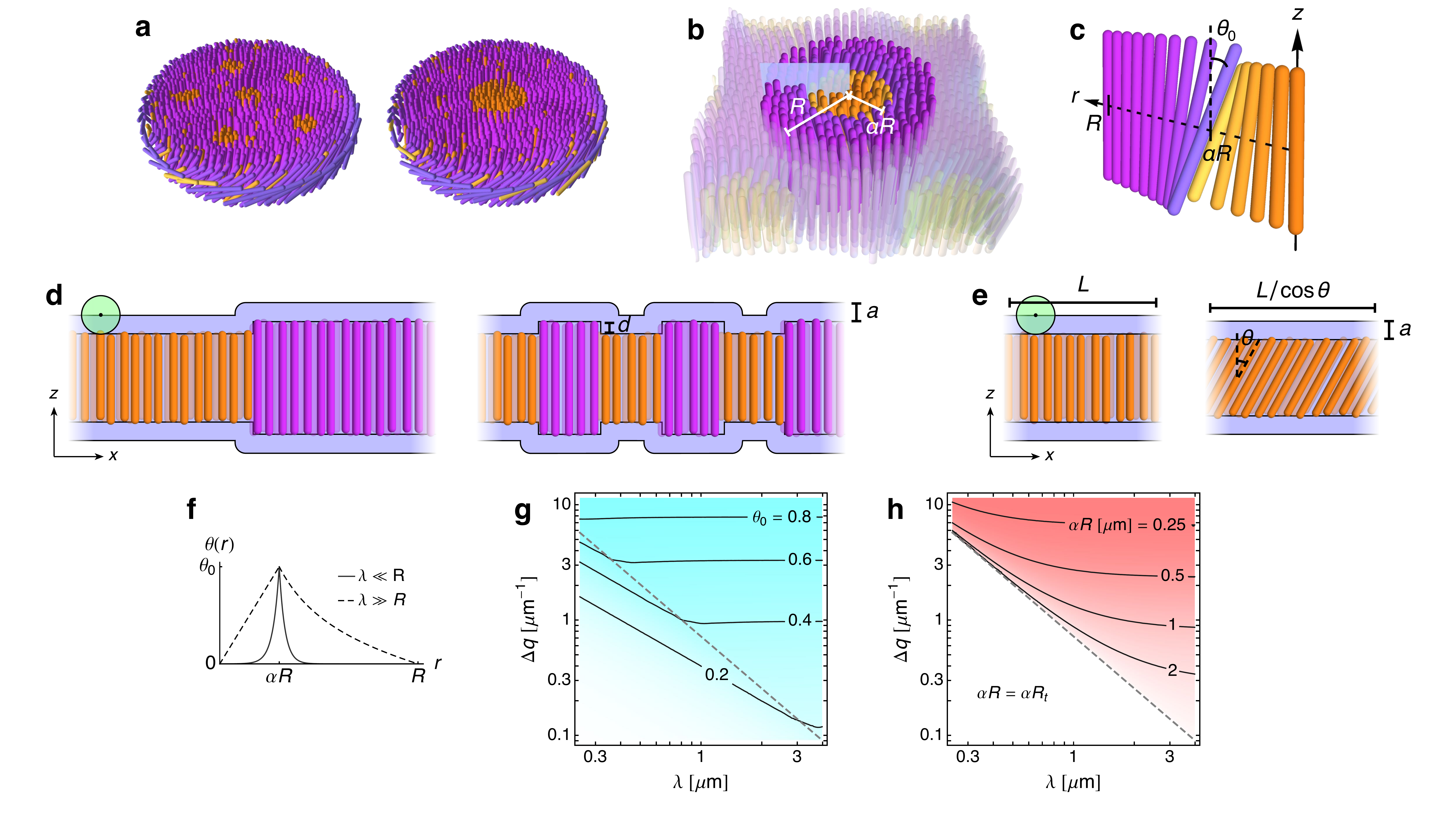}
	\caption{\label{fig:structure}Raft size and chiral structure. \textbf{a}, Schematics of two membranes with the same degree of phase separation and thus the same raft area fraction $\alpha^2$ containing either several smaller rafts (left) or one larger raft (right). \textbf{b}, A single circular domain with a single circular raft is repeated to approximately tile the membrane. \textbf{c}, Structure of the domain along the light blue plane in \textbf{b}. Along the radial coordinate $r$, the \fd viruses (orange) twist from $\theta(0) = 0$ to $\theta(\alpha R) = \theta_0$ at the raft-background interface with one handedness, and the background viruses, containing mostly \MK virus (purple), twist from $\theta(\alpha R) = \theta_0$ to $\theta(R) = 0$ at the domain edge with the other handedness. \textbf{d}--\textbf{e}, The effect of depletants (green circles) on raft structure and organization. \textbf{d}, Between two membranes of equal volume, the one with more interface between raft and background (right) has greater excluded volume (blue), leading to an interfacial line tension proportional to $d$. \textbf{e}, Between two membranes of equal volume, the one whose viruses are tilted at angle $\theta$ (right) has greater excluded volume, leading to a free energy term proportional to $\theta^2$ to leading order. \textbf{f}, Tilt angle $\theta(r)$ (Eq.~\ref{eqn:theta}) for domains whose common twist penetration depth $\lambda \equiv \lambda_1 \approx \lambda_2$ is much less or much greater than their radius $R$. \textbf{g}, Maximum twist angle $\theta_0$ (Eq.~\ref{eqn:theta0}) as a function of $\lambda$ and the twist wavenumber difference $\Delta q \equiv q_1 - q_2$. Darker cyan indicates larger $\theta_0$. \textbf{h}, Raft radius $\alpha R$ as a function of $\lambda$ and $\Delta q$, calculated numerically. Darker red indicates smaller $\alpha R$. We assume the large membrane limit $R_\rmt \rightarrow \infty$. The maximum raft radius $\alpha R_\rmt$ corresponds to a membrane having only a single raft, a regime separated by a gray dashed line from membranes with multiple smaller rafts (Eq.~\ref{eqn:critical}). This line is reproduced in \textbf{g}. For \textbf{g}--\textbf{h}, $\alpha = 0.3$ and values for other parameters are provided in Table~\ref{tab:raft}. Schematics not drawn to scale.}
\end{figure*}

Assuming we are in the regime $c d/c_\rmv > \log 1/\alpha_\rmt$ in which rafts exist, we now analyze their structure. Equation~\ref{eqn:alpha} determines the total amount of \fd virus sequestered into the raft phase by setting the value of $\alpha$, but does this phase form a single large raft or several smaller rafts (Fig.~\ref{fig:structure}a)? And how are the virus particles aligned? We will see that these questions are related via the natural tendency of chiral rods to adopt twisted configurations. To answer them, we need to derive the structural free energy of the membrane.

We take the membrane of radius $R_\rmt \rightarrow \infty$ to be approximately tiled by circularly symmetric domains of radius $R$ (Fig.~\ref{fig:structure}b), as in the muffin-tin approximation of solid state physics~\cite{Slater:1937un}. There are $R_\rmt^2/R^2$ domains and the total membrane free energy is
\begin{equation}
	F_\textrm{struct} = \frac{R_\rmt^2}{R^2} F_\textrm{domain},
\end{equation}
where $F_\textrm{domain}$ is the free energy of a single domain, which contains one raft of radius $\alpha R$. The \fd particles point vertically at the center of the raft and twist azimuthally with one handedness to their interface with the background, where they attain twist angle $\theta_0$. The background particles, which are mostly \MK with a smaller amount of \fd, twist with opposite handedness from $\theta_0$ at the interface to 0 at the domain edge, where the next domain would begin (Fig.~\ref{fig:structure}c). Once the membrane separates into its thermodynamically preferred raft and background phases, we assume zero net particle current between the phases and between each phase and the aqueous environment. We also assume that the 2D particle concentration $c_\rmv$ in the membrane is constant. Thus, the volume of each phase is conserved, so any effects of depletion must only act on the surface area of the membrane (Eq.~\ref{eqn:Fdepgen}). For mathematical tractability, we assume the particles do not twist very much, so their tilt angle satisfies $\theta \ll 1$, and the two virus species have similar half-lengths $l_\itfd$ and $l_\rmMK$, so their half-length difference satisfies $d \ll l_\itfd \approx l_\rmMK$. As calculated in Ref.~\cite{Kang:2016is}, virus position fluctuations perpendicular to the membrane are strongly suppressed in the $\theta \ll 1$ limit, so the thicknesses of the raft and background phases are simply $2l_\itfd \cos\theta$ and $2l_\rmMK \cos\theta$, respectively.

$F_\textrm{domain}$ consists of three components. First, interfaces between raft and background have a half-height difference of approximately $d$. These vertical offsets, which appear as ``corners'' in Fig.~\ref{fig:structure}d, contribute additional membrane surface area and, through the depletion free energy Eq.~\ref{eqn:Fdepgen}, produce an effective interfacial line tension proportional to $d$. Second, virus tilt away from the membrane normal also increases the membrane surface area and, also through depletion, produces an effective alignment energy proportional to $\theta^2$ (Fig.~\ref{fig:structure}e). Third, the virus particles behave as chiral nematic liquid crystals~\cite{Dogic:2000tp,Tombolato:2006hy}. That is, each species prefers to be aligned in a twisted configuration with wavenumber $q$, where the sign of $q$ indicates the chirality of twist (positive corresponds to right-handed) and $2\pi/|q|$ is the wavelength. The energetic cost of deviations from this preferred configuration is given by the Frank free energy~\cite{Frank:1958ey}:
\begin{align}
	F_\textrm{Frank} &= K \int \dd^2 \ve x\, l\cos\theta \nonum
	& \enspace\quad {}\times\left[(\bdiv\ve n)^2 + (\bcurl\ve n)^2 - 2 q \ve n\cdot\bcurl\ve n\right].
	\label{eqn:FFrankgen}
\end{align}
$\ve n$ is the nematic director, $K$ is the 3D Frank elastic constant in the one-constant approximation, $q$ is the preferred twist wavenumber associated with intrinsic chirality of the constituent particles, $l$ is the particle half-length, and $\theta$ is the particle tilt angle. For raft domains depicted in Fig.~\ref{fig:structure}b--c, the nematic director is circularly symmetric and tilts away from the membrane normal in the negative azimuthal direction:
\begin{equation}
	\ve n(r) = -\sin \theta(r)\, \veh\phi + \cos \theta(r)\, \veh z.
\end{equation}

\begin{widetext}
The complete derivation of $F_\textrm{domain}$ is given in Supporting Information, and it leads to the structural free energy
\begin{align}
	\frac{F_\textrm{struct}}{4\pi c aT} &= \frac{R_\rmt^2}{R^2} \Bigg\{d\alpha R - \left[\lambda_1^2 q_1 - \lambda_2^2 q_2 \right]\alpha R \theta_0 + \frac{1}{2}\left[\lambda_1^2 + \lambda_2^2\right]\theta_0^2 \nonum
	&\qquad\qquad {}+ \int_0^{\alpha R} \dd r\left[\frac{1}{2}r\theta^2 + \frac{\lambda_1^2}{2}\left(r(\partial_r\theta)^2 + \frac{\theta^2}{r}\right)\right] + \int_{\alpha R}^R \dd r\left[\frac{1}{2}r\theta^2 + \frac{\lambda_2^2}{2}\left(r(\partial_r\theta)^2 + \frac{\theta^2}{r}\right)\right]\Bigg\}.
	\label{eqn:F}
\end{align}
The subscripts 1 and 2 refer to raft and background phases respectively. An important lengthscale $\lambda_j \equiv \sqrt{K_jl_j/caT}$ arises from comparing the Frank twist and depletion contributions to the free energy, where $j \in \{1,2\}$. The latter penalizes nonzero $\theta(r)$ and the former penalizes gradients in $\theta(r)$, so $\lambda_j$ acts like a twist penetration lengthscale. Since only \fd viruses compose rafts, $q_1 = q_\itfd$ and $\lambda_1 = \sqrt{K_\itfd l_\itfd/caT}$.  The corresponding expressions for the background must account for a mixture of virus species. Experiments demonstrate that cholesteric mixtures of \itfd-wt and \fd viruses have intermediate twist wavenumbers that linearly interpolate between their pure values as a function of relative concentration~\cite{Barry:2009uv}. We assume that the same behavior applies here to Frank constants and twist wavenumbers for \fd and \MK viruses:
\begin{equation}
	q_2 = \frac{1-\alpha_\rmt^2}{1-\alpha^2} q_\rmMK + \frac{\alpha_\rmt^2-\alpha^2}{1-\alpha^2} q_\itfd \qquad\textrm{and}\qquad \lambda_2 = \sqrt\frac{K_2 l_\rmMK}{caT}, \quad\textrm{where}\quad K_2 = \frac{1-\alpha_\rmt^2}{1-\alpha^2}K_\rmMK + \frac{\alpha_\rmt^2-\alpha^2}{1-\alpha^2}K_\itfd.
\end{equation}
Experimental estimates for $K$ and virus half-length $l$ are of the same order of magnitude for the two species (Table~\ref{tab:raft}). For better mathematical insight and clearer presentation of results, we will sometimes imagine that they are equal, so the two phases share the same $\lambda \equiv \lambda_1 \approx \lambda_2$. Another important parameter is $\Delta q \equiv q_1 - q_2$, the difference between the chiral wavenumbers of the raft and background.

Minimization of $F_\textrm{struct}$ over the tilt angle $\theta(r)$ and the domain radius $R$ yields the thermodynamically preferred membrane structure. We first minimize over $\theta(r)$ with the boundary conditions $\theta(0) = 0$, $\theta(\alpha R) = \theta_0$, and $\theta(R) = 0$:
\begin{equation}
	\theta(r) = \begin{cases}\displaystyle \theta_0 \frac{I_1(s_1)}{I_1(\alpha S_1)} & 0 \leq r \leq \alpha R \\ \displaystyle \theta_0 \frac{K_1(s_2)/K_1(S_2) - I_1(s_2)/I_1(S_2)}{K_1(\alpha S_2)/K_1(S_2) - I_1(\alpha S_2)/I_1(S_2)} & \alpha R \leq r \leq R, \end{cases}
	\label{eqn:theta}
\end{equation}
where $I_\nu$ and $K_\nu$ are modified Bessel functions of the first and second kind, respectively, of order $\nu$ (the latter should not be confused for Frank constants). Distances are rescaled by the twist penetration depths as $s_j = r/\lambda_j$ and $S_j = R/\lambda_j$, for $j \in \{1, 2\}$. Solving the Euler-Lagrange equation is described in Supporting Information. Equation~\ref{eqn:theta} is plotted in Fig.~\ref{fig:structure}f. If the common twist penetration depth $\lambda$ is much less than $R$, then the twist is exponentially localized to the interface between raft and background, but if it is much greater than $R$, then the twist $\partial_r\theta$ extends uniformly throughout the membrane.

We then substitute Eq.~\ref{eqn:theta} into Eq.~\ref{eqn:F}, perform the integrals over $r$, and minimize over $\theta_0$, the tilt angle at the interface:
\begin{equation}
	\theta_0 = \frac{\lambda_1^2 q_1 - \lambda_2^2 q_2}{\lambda_1\frac{I_0(\alpha S_1)}{I_1(\alpha S_1)} + \lambda_2\frac{K_0(\alpha S_2)/K_1(S_2) + I_0(\alpha S_2)/I_1(S_2)}{K_1(\alpha S_2)/K_1(S_2) - I_1(\alpha S_2)/I_1(S_2)}}.
	\label{eqn:theta0}
\end{equation}
This equation is plotted in Fig.~\ref{fig:structure}g. The magnitude of $\theta_0$ increases with $\lambda$ and $\Delta q$, and its sign is determined by the sign of $\Delta q$.

Substituting Eq.~\ref{eqn:theta0} back into $F_\textrm{struct}$ yields
\begin{equation}
	\frac{F_\textrm{struct}}{4\pi R_\rmt^2 c a T} = \frac{\alpha}{R} \left\{d - \frac{1}{2}\frac{\left(\lambda_1^2 q_1 - \lambda_2^2 q_2\right)^2}{\lambda_1\frac{I_0(\alpha S_1)}{I_1(\alpha S_1)} + \lambda_2\frac{K_0(\alpha S_2)/K_1(S_2) + I_0(\alpha S_2)/I_1(S_2)}{K_1(\alpha S_2)/K_1(S_2) - I_1(\alpha S_2)/I_1(S_2)}}\right\},
	\label{eqn:Funsimp}
\end{equation}
which only depends on the free parameter $R$ through $S_1$ and $S_2$. By minimizing over $R$, we numerically calculate the preferred raft radius $\alpha R$, remembering that $\alpha$ was determined in the previous section. Figure~\ref{fig:structure}h shows that at low $\lambda$ and $\Delta q$, $R$ adopts its maximum value, $R_\rmt$, so the membrane contains one large raft. As $\Delta q$ increases past a critical value, $R$ prefers a finite value and the raft phase separates into several smaller rafts of radius $\alpha R$. For constant $\Delta q$, increasing $\lambda$---or equivalently decreasing $c$---leads to more numerous, smaller rafts, which qualitatively agrees with experimental observations in Fig.~\ref{fig:overview}c--d. Note that the chirality inversion $q_1 \rightarrow -q_1$ and $q_2 \rightarrow -q_2$ yields the mirror-image configuration $\theta(r) \rightarrow -\theta(r)$ via Eqs.~\ref{eqn:theta} and \ref{eqn:theta0} with same free energy Eq.~\ref{eqn:Funsimp}.
\end{widetext}

A large chiral twist wavenumber difference $\Delta q$ indicates the proclivity of \fd and \MK viruses to twist back and forth with opposite handednesses; however, depletants favor particle alignment perpendicular to the membrane. A large number of small rafts can satisfy both tendencies, since the particles can twist back and forth over short distances while largely maintaining perpendicular alignment. In opposition is the positive interfacial line tension also generated by depletion, which prefers a small number of large rafts in order to reduce the total interfacial length between raft and background phases. The competition between these factors sets the raft size, which we can see explicitly by expanding the free energy to leading orders in $R^{-1}$, corresponding to the phase transition between single- and multiple-raft membranes. With the simplification $\lambda \equiv \lambda_1 \approx \lambda_2$, Eq.~\ref{eqn:Funsimp} becomes
\begin{align}
	\frac{F_\textrm{struct}}{4\pi R_\rmt^2 c a T} &\sim \frac{\alpha}{R} \bigg\{d - \frac{1}{4}\lambda^3\Delta q^2 + \frac{3}{32} \frac{\lambda^5\Delta q^2}{\alpha^2 R^2} \nonum
	&\qquad\qquad\quad{}+ \frac{1}{4} \lambda^3\Delta q^2 \ee^{-2(1-\alpha)R/\lambda} \bigg\}.
	\label{eqn:FK}
\end{align}
Thus, virus chirality appends a correction term to the bare interfacial tension to produce the effective interfacial line tension $2caT(d-\lambda^3\Delta q^2/4)$. When this effective tension becomes negative, the system prefers multiple smaller rafts instead of a single large raft in order to increase the total interfacial length. The critical dashed line of Fig.~\ref{fig:structure}h occurs when it equals zero and is thus given by
\begin{equation}
	|\Delta q| = 2 d^{1/2}\lambda^{-3/2}.
	\label{eqn:critical}
\end{equation}
In the multiple-raft regime where $|\Delta q|$ exceeds this critical value, the preferred raft size is
\begin{equation}
	\alpha R \sim \sqrt{\frac{\frac{9}{32}\lambda^5\Delta q^2}{\frac{1}{4}\lambda^3\Delta q^2 - d}},
	\label{eqn:alphaR}
\end{equation}
indicating a second-order phase transition. Notice that Eq.~\ref{eqn:FK} is analogous to the free energy of the 2D Frenkel-Kontorova model around the commensurate-incommensurate transition, with the first two terms corresponding to an effective interfacial line tension between rafts and background, the third corresponding to what can be interpreted as an effective interfacial bending energy, and the fourth corresponding to raft-raft repulsion~\cite{Bak:1982it,chaikinlubensky}. The higher-order terms prevent a negative effective interfacial tension from decreasing the raft size to 0 and set the preferred size Eq.~\ref{eqn:alphaR}.

\begin{figure}
	\includegraphics[width=\columnwidth]{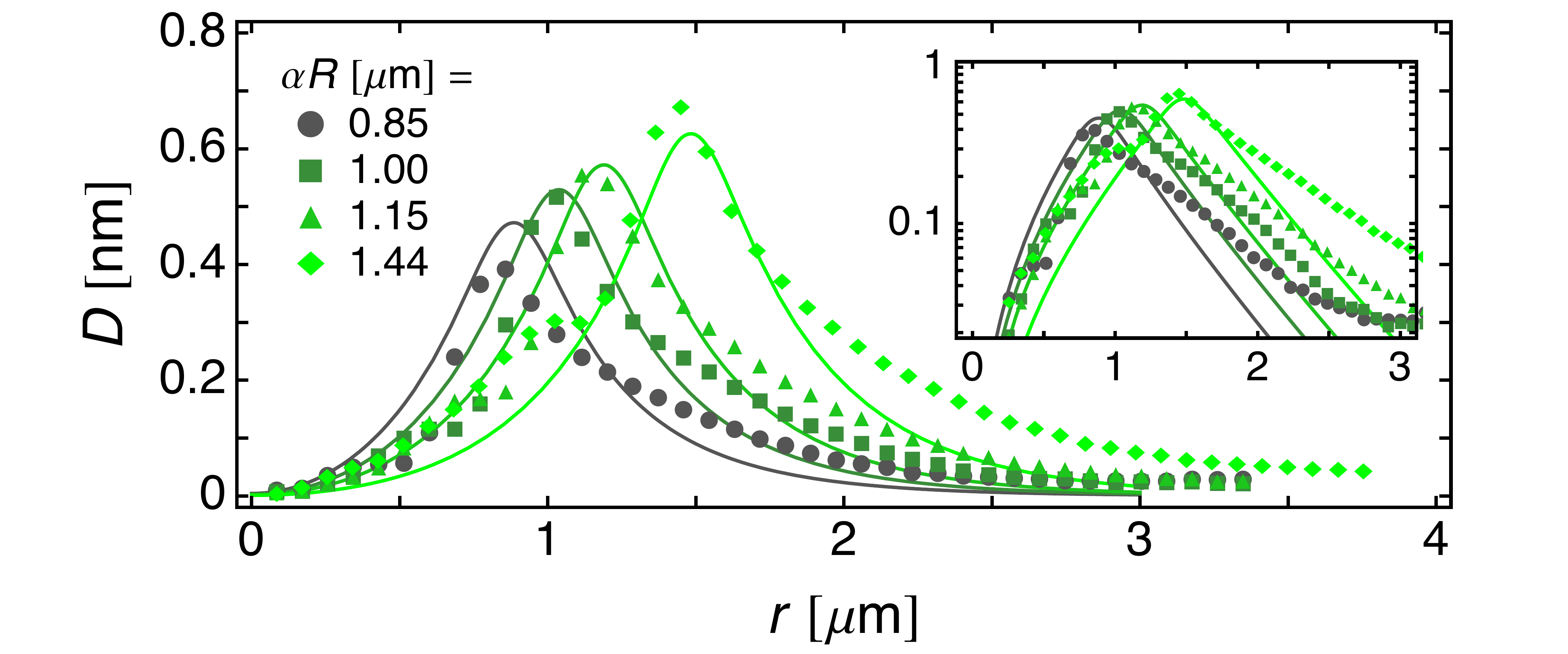}
	\caption{\label{fig:ret}Retardance values $D$ for rafts of various radii $\alpha R$. The points indicate experimental data and the lines indicate theoretical results calculated with $\alpha_\rmt = 0.5$ and the parameter values in Table~\ref{tab:raft}, corresponding to twist penetration depth $\lambda \sim \SI{0.8}{\um}$ and chiral wavenumber difference $\Delta q = \SI{0.5}{\per\um}$. $\alpha$ is given by Eq.~\ref{eqn:alpha} and $R$ is adjusted to produce rafts of different radii. Experimental data and methods are reported in Ref.~\cite{Sharma:2014cl}.}
\end{figure}

To assess the validity of our model, we can compare measurements of optical retardance (Fig.~\ref{fig:overview}f) to values calculated by our model. When polarized light passes through a birefringent material, the ``ordinary'' and ``extraordinary'' components propagate at different speeds, leading to a phase difference called retardance that we measure in wavelengths. For our membranes, it is approximately given by $D = 2\Delta n l \sin^2\theta$ and is thus an indirect measure of the tilt angle $\theta$~\cite{bornwolf}. The raw calculated retardance profiles are convolved with a Gaussian of width \SI{0.13}{\um} representing the microscope's resolution function, exactly as previously reported~\cite{Barry:2009vo}. Figure~\ref{fig:ret} shows good agreement between theoretical and experimental retardance profiles using the physically reasonable birefringence values reported in Table~\ref{tab:raft}.

\section{Raft-raft repulsion}

\begin{figure}
	\includegraphics[width=\columnwidth]{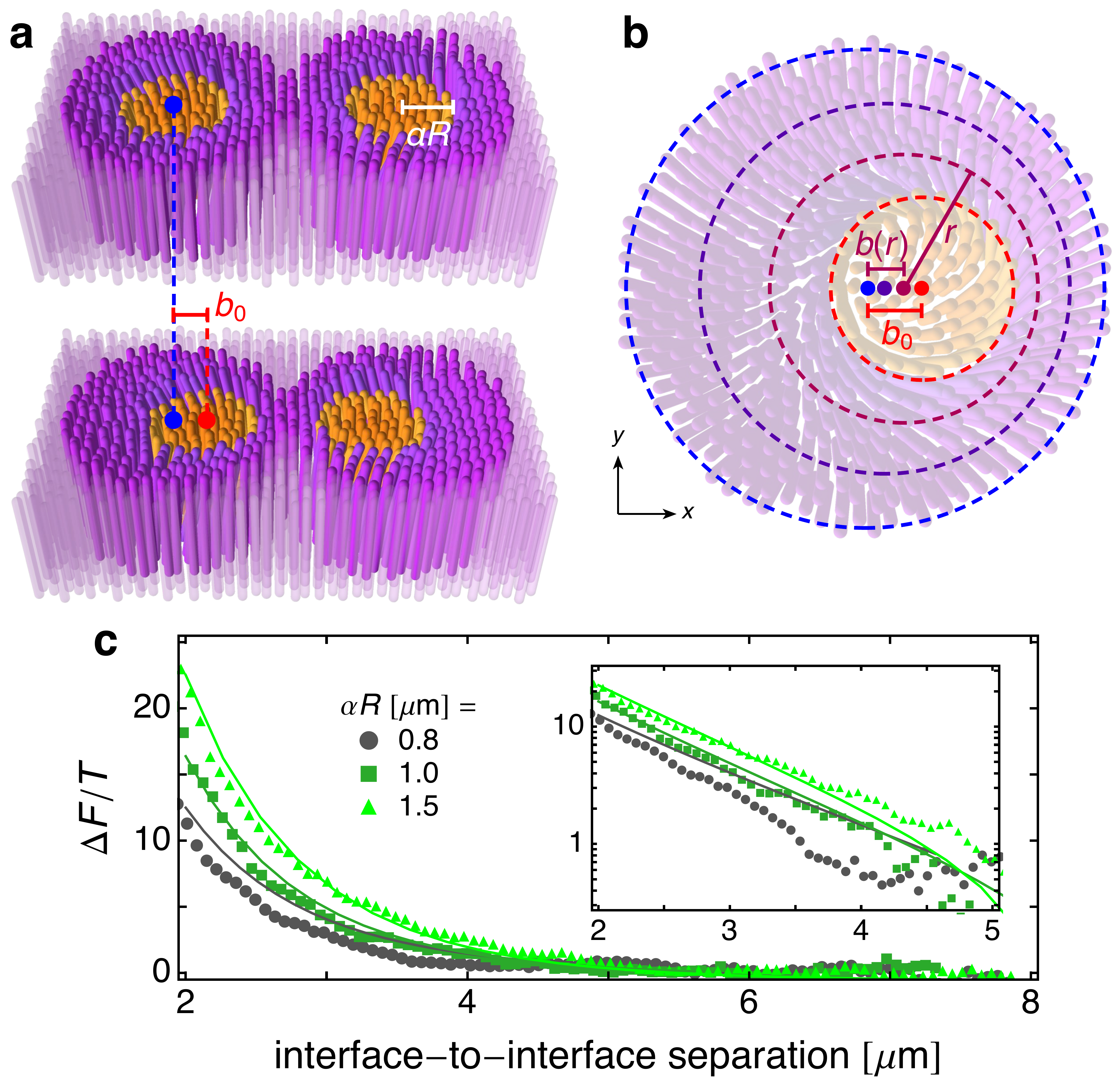}
	\caption{\label{fig:rep}Raft-raft repulsion. \textbf{a}, The approach of two rafts is modeled as raft shifts $b_0$ with respect to their circular tiling domains. \textbf{b}, Shifted polar coordinate system of the background membrane (Eq.~\ref{eqn:xy}). Dashed lines indicate curves of constant $r$ from $r = \alpha R$ (red) to $r = R$ (blue), which are circles of radius $r$ whose centers (dots) lie at $x = b(r)$ and $y = 0$. \textbf{c}, Raft-raft repulsion energy $\Delta F$ divided by temperature $T$ for rafts of various radii $\alpha R$. The points indicate experimental data and the lines indicate theoretical results calculated with $\alpha_\rmt = 0.5$ and the parameter values in Table~\ref{tab:raft}, corresponding to twist penetration depth $\lambda \sim \SI{0.8}{\um}$ and chiral wavenumber difference $\Delta q = \SI{0.5}{\per\um}$. $\alpha$ is given by Eq.~\ref{eqn:alpha} and $R$ is adjusted to produce rafts of different radii. Experimental data and methods are reported in Ref.~\cite{Sharma:2014cl}. Schematics not drawn to scale.}
\end{figure}

To model the interaction between two neighboring rafts as they approach each other, we shift each circular raft within its circular tiling domain off-center by a distance $b_0$ towards each other (Fig.~\ref{fig:rep}a). To accomplish this, the background membrane must be deformed; for simplicity, we assume that the rafts themselves are unchanged by this shift. We parametrize the deformation by a shift profile $b(r)$ such that Cartesian coordinates are given in terms of shifted polar coordinates by
\begin{equation}
	x = r \cos\phi + b(r) \qquad\textrm{and}\qquad y = r \sin\phi.
	\label{eqn:xy}
\end{equation}
In other words, the curves of constant $r$ are nested non-concentric circles of radius $r$ centered at $x = b(r)$ and $y = 0$ (Fig.~\ref{fig:rep}b). Our shift ansatz Eq.~\ref{eqn:xy} breaks circular symmetry into dipolar symmetry, implying that $\theta$ can vary with azimuthal angle $\phi$ and that particles can tilt in the $\veh r$ direction. To dipolar order, $\theta(r,\phi) = \theta(r) + \vartheta(r)\cos\phi$, where $\vartheta(r)$ is the dipolar tilt component. We must carefully recalculate terms in the single-domain free energy $F_\textrm{domain}$ that would be changed by this coordinate transformation:
\begin{align}
	\frac{F_\textrm{shift}}{2 c aT} &= \int_{\alpha R}^R \dd r\int_0^{2\pi} \dd\phi\, h_r h_\phi \nonum
	& \qquad{}\times \left\{\frac{1}{2} \theta^2 + \frac{\lambda_2^2}{2} \left[(\bdiv \ve n)^2 + (\bcurl \ve n)^2\right]\right\}.
	\label{eqn:Fshiftgen}
\end{align}
$h_r$ and $h_\phi$ are scale factors of the coordinate transformation. The evaluation of Eq.~\ref{eqn:Fshiftgen} is provided in Supporting Information, where we see that the $\veh r$ component of $\ve n$ can be ignored to leading order in tilt angles. The shift profile $b(r)$ appears from the scale factors and the spatial derivatives. Since we assume the rafts are unchanged by the deformation, $\theta(\alpha R) = \theta_0$ and $\vartheta(\alpha R) = 0$, corresponding to the unshifted interfacial tilt angle as given by Eq.~\ref{eqn:theta0}. The deformation vanishes at the edge of the tiling domain, so $b(R) = 0$, $\theta(R) = 0$, and $\vartheta(R) = 0$. To calculate the interaction energy between two rafts as a function of separation distance, we impose various shift distances $b_0 = b(\alpha R)$; numerically minimize the free energy over $b(r)$, $\theta(r)$, and $\vartheta(r)$; subtract the energy of the unshifted membrane with $b(r) = 0$; and double the result.

Meanwhile, the repulsive free energy of this two-raft system has been measured experimentally via optical trapping by moving rafts toward each other, releasing them, and tracking their subsequent trajectories (Fig.~\ref{fig:overview}g and \cite{Sharma:2014cl}). Using parameter values given in Table~\ref{tab:raft}, our model agrees well with these measurements for various raft radii $\alpha R$ (Fig.~\ref{fig:rep}c). Thus, despite our relatively simple ansatz, our results quantitatively demonstrate that deformation of the background membrane as two rafts approach each other can explain the observed repulsion between rafts.

\section{Discussion}

Our model is designed to emphasize physical relevance and minimize phenomenological contributions. To do so, we ignore many effects that may ultimately produce a more precise description of these colloidal membranes, but in the process add more fit parameters that obscure the underlying generalizable physical principles. For example, the viruses are idealized to be hard rods that form geometrically precise and homogeneous membranes. During phase separation, we disregard the increased translational entropy of the shorter \fd viruses when they are embedded within the longer \MK viruses. Furthermore, for mathematical tractability, we expand the membrane free energy to quadratic order in $d/l_j$ and $\theta$, even though the values in Table~\ref{tab:raft} imply $d/l_\itfd = 0.3$ and $\theta_0 \approx 0.25$.

On the other hand, our conceptual division of raft formation into the two sequential steps of phase separation and raft organization appears to be justified. Numerical minimization of a free energy combining Eqs.~\ref{eqn:Fsep} and \ref{eqn:Funsimp} yields results indistinguishable from Figs.~\ref{fig:mixing}d and \ref{fig:structure}h, indicating that the characteristic energy scale of phase separation is much higher than that of raft organization (Supporting Information). Moreover, this division is demonstrated in the experimental separation of relaxation timescales. As depicted in Fig.~\ref{fig:overview}e, rafts take ${\sim}\SI{24}{\hour}$ to reach their equilibrium size, but the membrane reaches its equilibrium degree of phase separation much more quickly (the background fluorescence stays constant throughout the three panels). Both processes undergo energetic relaxation through diffusion of the same particles, so their decay timescales scale as $\tau \sim \eta/\varepsilon$, where $\eta$ is the viscosity and $\varepsilon$ is an energy density scale. A larger $\tau$ for the process of raft organization corresponds to a smaller $\varepsilon$ compared to that of membrane phase separation, which our model explains.

Despite these sweeping simplifications, our model can match measurements with quantitative accuracy while using physically reasonable parameter values. It is consistent with our single-component membrane model that described an independent set of experimental observations~\cite{Kang:2016is}. Moreover, it provides meaningful insight into the fundamental mechanisms that drive membrane raft formation and organization. Competition between mixing entropy and depletion entropy determines the degree of phase separation of two virus species with different lengths. This competition is independent of virus chirality can be easily and precisely tuned by adjusting the depletant concentration. A difference in the natural tendency for chiral particles to twist with a preferred handedness and pitch endows the rafts with a chiral structure. This structure stabilizes small rafts against an interfacial line tension that would otherwise promote coarsening to a single raft domain and establishes a preferred depletant-concentration-dependent raft size. The twisted structure of the background membrane transmits torques and mediates an elastic repulsion between rafts.

Previous theoretical reports have demonstrated that chiral structure can establish a membrane lengthscale, but they differ from our theory in several crucial ways. Some describe single-component smectic-\textit{C} membranes that contain hexagonal cells with only one handedness of twist and arrays of defects at the corners of the cells~\cite{Hinshaw:1988gb,Hinshaw:1989bd}.  Selinger and colleagues investigate membranes formed from racemic mixtures that can form domains of alternating chirality upon spontaneous symmetry breaking~\cite{Selinger:1993bm}. They find a square lattice of domains that also contain defects at their corners. Simultaneously with our work, their theory has been expanded to hexagonal domains without defects and applied to filamentous virus membranes~\cite{sakhardande}. These aforementioned theories are based on phenomenological Landau expansions in the concentration difference between the two chiral components (we show how our model can provide values for Landau coefficients in Supporting Information). Complementarily, Xie and colleages investigate raft-raft repulsion by directly minimizing the free energy of both raft and background~\cite{Xie:2016gv}. They highlight the role of background chiral twist and use values for Frank constants ($\SI{5}{\pico\newton}$) and twist wavenumbers (${\sim}\SI{3}{\per\um}$) that are within an order of magnitude of those we use (Table~\ref{tab:raft}). However, they assume well-defined rafts of a particular size, ignoring the processes of phase separation and raft size establishment, and use a phenomenological virus tilt modulus without exploring its physical basis in depletion entropy. In contrast, our theory, which provides a more unified microscopic approach that facilitates comparison with experiments, produces analytical expressions for the chiral raft structure, and provides mathematical intuition for raft-raft repulsion via a shift ansatz.

Colloidal membranes composed of viruses share important physical symmetries with their molecular counterparts, even though their characteristic lengthscales and microscopic origins of interactions differ. In fact, a leading-order free energy for rafts in a flat molecular membrane would look very much like Eq.~\ref{eqn:F}. The interfacial line tension between rafts and background would replace the term proportional to $d$~\cite{Samsonov:2001gl,Kuzmin:2005hv}. Phases that prefer alignment perpendicular to the membrane plane, such as the biologically-relevant L$_\alpha$ phase, would require a $\theta^2$ term~\cite{katsarasgutberlet,Tardieu:1973fs}. Molecular twist would be encapsulated by Frank free energy terms. A generalization of our model which can be applied to other membrane systems is provided in Supporting Information. Furthermore, experimentally-prepared and biological membranes have rafts enriched in cholesterol as compared to the background~\cite{Simons:1997jq,alberts}. Cholesterol demonstrates a strong preference for chiral twist---in fact, the chiral nematic, or cholesteric, phase was the first liquid crystalline phase observed in 1888 by Friedrich Reinitzer while investigating cholesteryl esters~\cite{Reinitzer:1888kw}. Hence, we expect a significant difference in chiral wavenumbers $\Delta q$ which could stabilize smaller rafts.

Our theory contributes to a biologically-relevant and poorly-understood niche in the rich literature on molecular membranes. It may explain why Langmuir monolayers composed of multiple chiral molecules demonstrate a limit to domain coarsening~\cite{Seul:1994gn} and biological lipid rafts are believed to have a finite size~\cite{Anderson:2002bb}, in contradiction to continous coarsening predicted by the Cahn-Hilliard model of phase separation~\cite{chaikinlubensky}. Our description of raft-raft repulsion is analogous to the twist-mediated interaction of chiral islands in smectic-\textit{C} films~\cite{Cluzeau:2002dt,Bohley:2008dc,Silvestre:2009jg}. It offers an explanation for the mutual repulsion observed between transmembrane protein pores formed by certain antimicrobials, if one imagines that these chiral pores impose phospholipid tilt at their interface with the background membrane~\cite{Constantin:2007ww,Constantin:2009iu}. Ultimately, the validity of our theory in a particular membrane system hinges on the direct observation of twist, which can be achieved with polarized optical microscopy if the twist penetration depth is at least the wavelength of light~\cite{Gibaud:2012cf}.

Moreover, phospholipid rafts demonstrate chiral phase behavior that must be explained by a theory attuned to chirality. By either replacing naturally chiral sphingomyelins with a racemic mixture~\cite{Ramstedt:1999dx} or replacing cholesterol with its enantiomer~\cite{Lalitha:2001br,Lalitha:2001bu} (although these latter studies disagree with subsequent work~\cite{Westover:2003di,Westover:2004jf}), the critical point for phase separation changes. Although for our model parameters, phase separation occurs independently from raft organization, other parameter values cause the raft area fraction $\alpha^2$ to depend on the difference in chiral twist wavenumbers $\Delta q$ (Supporting Information). Furthermore, different enantiomers of the same anesthetic molecule have been shown to have different potencies~\cite{Lysko:1994wd,Dickinson:2000ww,Won:2006fs}. Our theory presents a paradigm through which chirality affects physical membrane properties, in accordance with the classic hypothesis that anesthetic molecules disrupt membrane phase behavior~\cite{Campagna:2003gd,Weinrich:2012cr}.

\begin{acknowledgments}
We are grateful to Zvonimir Dogic and Prerna Sharma for generously suggesting ideas, sharing data, and critically reading our manuscript. We thank our reviewers for offering constructive suggestions. We acknowledge financial support from the National Science Foundation through grant DMR-1104707. T.C.L. is grateful for support from a Simons Investigator grant.
\end{acknowledgments}

\bibliography{refs,refs-book}

\end{document}